\documentclass{emulateapj}
\usepackage{apjfonts}

\newcommand{\pivec}{\mbox{\boldmath $\pi$}}

\lefthead{JUNG ET AL.}
\righthead{ }

\begin{document}
\title{OGLE-2013-BLG-0102LA,B: MICROLENSING BINARY WITH COMPONENTS AT 
STAR/BROWN-DWARF AND BROWN-DWARF/PLANET BOUNDARIES}

\author{
Y. K.~Jung$^{1}$, 
A.~Udalski$^{O1,O}$,
T.~Sumi$^{M1,M}$,
C.~Han$^{1,U,\dag}$,
A.~Gould$^{U1,U}$, \\
and\\
J.~Skowron$^{O1}$,
S.~Koz{\l}owski$^{O1}$,
R.~Poleski$^{O1,U1}$,
\L.~Wyrzykowski$^{O1,O2}$,
M.~K.~Szyma\'nski$^{O1}$,
G.~Pietrzy\'nski$^{O1,O3}$,
I.~Soszy\'nski$^{O1}$,
K.~Ulaczyk$^{O1}$,
P.~Pietrukowicz$^{O1}$,
P.~Mr{\'o}z$^{O1}$,
M.~Kubiak$^{O1}$ \\
(The OGLE Collaboration), \\
F.~Abe$^{M2}$, 
D.~P.~Bennett$^{M3}$,
I.~A.~Bond$^{M4}$, 
C.~S.~Botzler$^{M5}$, 
M.~Freeman$^{M5}$,
A.~Fukui$^{M6}$,
D.~Fukunaga$^{M2}$,
Y.~Itow$^{M2}$,
N.~Koshimoto$^{M1}$,
P.~Larsen$^{M7}$,
C.~H.~Ling$^{M4}$,
K.~Masuda$^{M2}$,
Y.~Matsubara$^{M2}$,
Y.~Muraki$^{M2}$,
S.~Namba$^{M1}$,
K.~Ohnishi$^{M8}$,
L.~Philpott$^{M9}$,
N.~J.~Rattenbury$^{M5}$,
To.~Saito$^{M10}$,
D.~J.~Sullivan$^{M11}$,
D.~Suzuki$^{M1}$, 
P.~J.~Tristram$^{M12}$, 
N.~Tsurumi$^{M2}$,
K.~Wada$^{M1}$,
N.~Yamai$^{M13}$,
P.~C.~M.~Yock$^{M5}$,
A.~Yonehara$^{M13}$ \\
(The MOA Collaboration), \\
M.~Albrow$^{U2}$,
J.-Y.~Choi$^{1}$, 
D.~L.~DePoy$^{U3}$,
B.~S.~Gaudi$^{U1}$,
K.-H.~Hwang$^{1}$,
C.-U.~Lee$^{U4}$,
H.~Park$^{1}$,
S.~Owen$^{U1}$,
R.~W.~Pogge$^{U1}$,
I.-G.~Shin$^{1}$,
J.~C.~Yee$^{U1,U5}$\\
(The $\mu$FUN Collaboration) \\
}

\bigskip\bigskip
\affil{$^{1}$Department of Physics, Chungbuk National University, Cheongju 371-763, Republic of Korea}
\affil{$^{O1}$Warsaw University Observatory, Al. Ujazdowskie 4, 00-478 Warszawa, Poland}
\affil{$^{O2}$Institute of Astronomy, University of Cambridge, Madingley Road, Cambridge CB3 0HA, UK}
\affil{$^{O3}$Universidad de Concepci\'on, Departamento de Astronomia, Casilla 160-C, Concepci\'on, Chile}
\affil{$^{M1}$Department of Earth and Space Science, Osaka University, Osaka 560-0043, Japan}
\affil{$^{M2}$Solar-Terrestrial Environment Laboratory, Nagoya University, Nagoya, 464-8601, Japan}  
\affil{$^{M3}$Department of Physics, University of Notre Dame, 225 Nieuwland Science Hall, Notre Dame, IN 46556-5670, USA}
\affil{$^{M4}$Institute of Information and Mathematical Sciences, Massey University, Private Bag 102-904, North Shore Mail Centre, Auckland, New Zealand}
\affil{$^{M5}$Department of Physics, University of Auckland, Private Bag 92-019, Auckland 1001, New Zealand}
\affil{$^{M6}$Okayama Astrophysical Observatory, National Astronomical Observatory of Japan, Asakuchi, Okayama 719-0232, Japan}
\affil{$^{M7}$Institute of Astronomy, University of Cambridge, Madingley Road, Cambridge CB3 0HA, UK}
\affil{$^{M8}$Nagano National College of Technology, Nagano 381-8550, Japan}
\affil{$^{M9}$Department of Physics and Astronomy, The University of British Columbia, 6224 Agricultural Road Vancouver, BC V6T 1Z1, Canada} 
\affil{$^{M10}$Tokyo Metropolitan College of Aeronautics, Tokyo 116-8523, Japan}
\affil{$^{M11}$School of Chemical and Physical Sciences, Victoria University, Wellington, New Zealand}
\affil{$^{M12}$Mt. John University Observatory, P.O. Box 56, Lake Tekapo 8770, New Zealand}
\affil{$^{M13}$Department of Physics, Faculty of Science, Kyoto Sangyo University, 603-8555, Kyoto, Japan}
\affil{$^{U1}$Department of Astronomy, Ohio State University, 140 West 18th Avenue, Columbus, OH 43210, USA}
\affil{$^{U2}$Department of Physics and Astronomy, University of Canterbury, Private Bag 4800, Christchurch, New Zealand}
\affil{$^{U3}$Department of Physics and Astronomy, Texas A\&M University, College Station, Texas 77843-4242, USA}
\affil{$^{U4}$Korea Astronomy and Space Science Institute, 776 Daedukdae-ro, Daejeon, Korea}
\affil{$^{U5}$Harvard-Smithsonian Center for Astrophysics, 60 Garden St., Cambridge, MA 02138, USA}
\affil{$^{O}$Optical Gravitational Lensing Experiment (OGLE) Collaboration}
\affil{$^{M}$Microlensing Observations in Astrophysics (MOA) Collaboration}
\affil{$^{U}$Microlensing Follow Up Network ($\mu$FUN) Collaboration}
\affil{$^{\dag}$Corresponding author}

\begin{abstract}
We present the analysis of the gravitational microlensing event OGLE-2013-BLG-0102. 
The light curve of the event is characterized by a strong short-term anomaly superposed on 
a smoothly varying lensing curve with a moderate magnification $A_{\rm max}\sim 1.5$. 
It is found that the event was produced by a binary lens with a mass 
ratio between the components of $q = 0.13$ and the anomaly was caused by the passage 
of the source trajectory over a caustic located away from the barycenter of the binary. 
From the analysis of the effects on the light curve due to the finite size of the source 
and the parallactic motion of the Earth, the physical parameters of the lens system 
are determined. The measured masses of the lens components are 
$M_{1} = 0.096 \pm 0.013~M_{\odot}$ and $M_{2} = 0.012 \pm 0.002~M_{\odot}$, 
which correspond to near the hydrogen-burning and deuterium-burning mass limits, respectively.
The distance to the lens is $3.04 \pm 0.31~{\rm kpc}$ and the projected separation 
between the lens components is $0.80 \pm 0.08~{\rm AU}$. 
\end{abstract}

\keywords{binaries: general, brown dwarfs -- gravitational lensing: micro}

\section{Introduction}

\begin{figure*}[th]
\epsscale{0.8}
\plotone{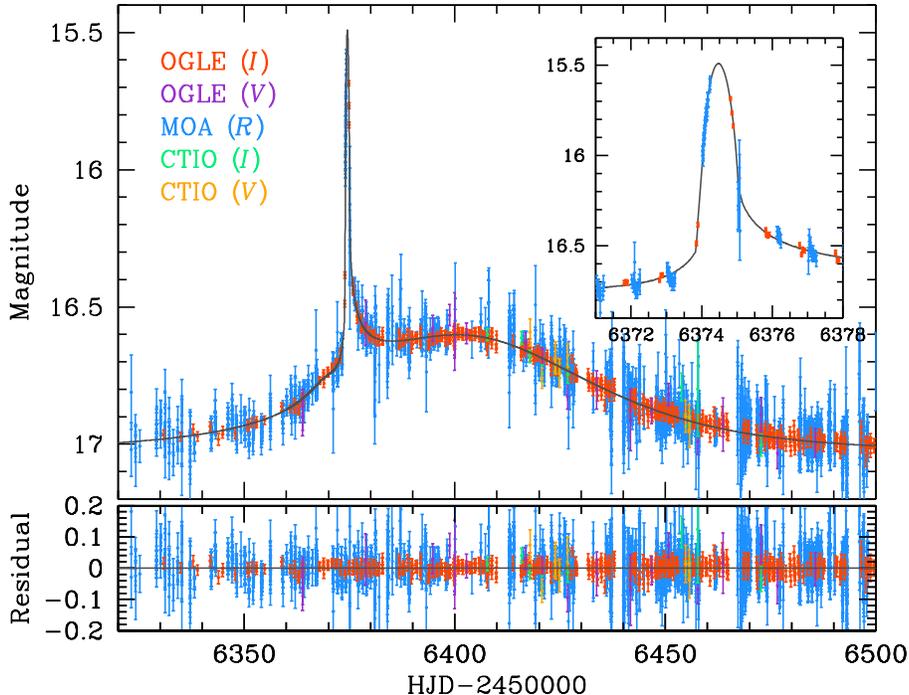}
\caption{\label{fig:one}
Light curve of OGLE-2013-BLG-0102. 
The inset shows the enlarged view of the anomaly centered at 
${\rm HJD}'= {\rm HJD}-2450000 \sim 6374.5$. 
The curve superposed on the data is the best-fit model. 
}
\end{figure*}

It is generally agreed that stars form through the collapse of 
gas clouds in interstellar medium while planets form either 
through coagulation of dust in protostellar 
disks or gravitational instabilities of the gas disk.
By contrast, there exist diverse mechanisms proposed to explain
the formation of objects with masses between stars and planets, 
i.e.\ very low-mass stars and brown dwarfs,including 
dynamical interaction \citep[e.g.,][]{boss01,reipurth01,umbreit05,bate09}, 
photoionizing radiation \citep[e.g.,][]{hester96,whitworth04}, 
disk instability \citep[e.g.,][]{goodwin07,stamatellos07}, 
turbulent fragmentation \citep[e.g.,][]{padoan04,hennebelle08}, etc.  
See \citet{luhman12} for detailed review.

In order to test formation theories of very low-mass (VLM) objects, 
various types of observations are needed. In this sense, 
observational studies of binaries composed of VLM objects are important 
because predictions of binary properties vary considerably among formation 
theories. For example, the dynamical interaction and the disk instability 
mechanisms predict few widely separated binaries 
while other mechanisms make no concrete predictions on such a trend.
Furthermore, with the exception of extreme microlensing events 
\citep{gould97, gould09} and astrometric microlensing \citep{cushing14} 
binaries provide the only channel to measure model-independent 
physical parameters including masses.

Unfortunately, comprehensive studies of VLM binaries have been difficult 
due to the lack of unbiased samples. Most known low-mass binaries 
have been discovered by direct imaging \citep[e.g.][]{close03, close07}. 
Due to the intrinsic nature, faint or dark VLM objects 
cannot be seen by this method and thus the sample favors binaries with luminous components 
and roughly equal masses, although some binaries found by other methods such as the 
``astrometric variable'' \citep[e.g.][]{dahn08, dupuy12, sahlmann13} and ``blended light 
spectroscopy'' \citep[e.g.][]{burgasser10, bardalez14} have little or no such bias.  
In addition, since it is difficult to detect closely separated ($\lesssim 1$ AU) binary 
systems due to the limitation set by angular resolution, the sample is biased toward 
widely separated binaries.  For the same reason, the sample is confined to binaries in 
the Solar neighborhood.  Furthermore, spectroscopic radial-velocity observations for 
these objects are difficult due to their faintness and thus it is difficult to precisely 
measure their masses.

Because of the difference in sensitivity from other methods, gravitational microlensing 
provides a complementary tool to study VLM objects.  Microlensing occurs due to the 
bending of light caused by the gravity of a lensing object located between an observer 
and a lensed star (source).  As a result, the phenomenon does not depend on the brightness 
of lensing objects, making it possible to detect faint and even dark objects.  Lensing 
events occur in Galactic scale and thus the method can be used to detect VLM binaries distributed 
over a wide range of Galactocentric distances.  In addition, the method is sensitive to 
tight binaries with small separations \citep{choi13}.  Furthermore, for well observed 
binary-lens events, it is possible to precisely measure binary masses without 
additional follow-up observations.  The method already demonstrated the usefulness in 
detecting VLM objects existing in various forms, e.g. a free-floating brown dwarf 
\citep{gould09}, binary brown dwarfs \citep{choi13}, brown dwarfs around stars 
\citep{shin12,street13}, and a brown dwarf orbited by a planetary mass object \citep{han13}.

In this paper, we report a low-mass binary discovered from the observation of the 
microlensing event OGLE-2013-BLG-0102.  We demonstrate that the binary is composed of 
a primary near the hydrogen-burning limit and a companion near the deuterium-burning limit.

\section{Observation}

The lensing event OGLE-2013-BLG-0102 occurred on a star located 
toward the Galactic bulge direction with the equatorial coordinates 
$(\alpha,\delta)_{\rm J2000}
=(17^{\rm h}52^{\rm m}07^{\rm s}\hskip-2pt.08, -31^{\circ}41'26''\hskip-2pt.1)$, 
corresponding to the Galactic coordinates $(l,b)=(358.36^\circ, -2.626^\circ)$. 
The event was first discovered by the Optical Gravitational Lensing Experiment 
\citep[OGLE:][]{udalski03} collaboration from the survey conducted 
toward the Galactic bulge field using the 1.3m 
Warsaw telescope at Las Campanas Observatory in Chile 
and the discovery was announced to the microlensing community on March 2, 2013. 
The event was independently discovered by the Microlensing Observations in Astrophysics 
\citep[MOA:][]{bond01,sumi03} collaboration using the 1.8m telescope at
Mt.~John Observatory in New Zealand and dubbed as MOA-2013-BLG-127. 
The MOA group noticed that the event had undergone an anomaly 
and issued a further alert for follow-up observations. 
No immediate follow-up observation could be done 
because the anomaly occurred during the very early Bulge season 
when the duration of the bulge visibility was short and 
telescopes for follow-up observations were not fully operational. 
Fortunately, the cadence of the survey observations was high enough to 
delineate the anomaly covering both the rising and falling parts of the anomaly.

From modeling of the light curve conducted by the time that the anomaly 
ended, it was suggested that the anomaly was produced by the crossing of a 
caustic\footnote[1]{The caustic represents the closed curve of formally 
infinite magnification on the source plane.} 
formed by a binary lens where the mass ratio between the components is low. 
In response to the potential importance of the event, 
the Microlensing Follow-Up Network 
\citep[$\mu$FUN:][]{gould06} collaboration took multiband images 
(14 images in $I$ band and 12 images in $V$ band) 
during April 25 -- June 29 period using the 1.3m SMARTS telescope of 
Cerro Tololo Inter-American Observatory (CTIO) in Chile 
to obtain the color information of the source star. 
The event lasted more than 100 days after the anomaly. 
From multiple stage real-time modeling of the light curve 
conducted with the progress of the event, 
it was suggested that higher-order effects would be needed to 
precisely describe the event.

Figure~\ref{fig:one} shows the light curve of the event. 
It is characterized by a strong short-term anomaly 
centered at ${\rm HJD}'= {\rm HJD}-2450000 \sim 6374.5$ 
superposed on a smooth brightness variation of the source star. 
The event lasted throughout the whole Bulge season and 
the anomaly lasted $\sim$ 4 days.

Data sets used for analysis were reduced using photometry codes developed 
by the individual groups, which are based on difference image analysis \citep{alard98}. 
For the use of data sets processed by different photometry systems, 
we readjust error bars. In this process, we first add a quadratic term so that 
the cumulative distribution of $\chi^{2}$ sorted by magnification 
becomes approximately linear. We then rescale the error bars so that $\chi^{2}$ 
per degree of freedom $(\chi^{2}/{\rm dof})$ for each data set becomes unity \citep{dong07}. 
The first process is needed to ensure that the dispersion of data points is 
consistent with error bars regardless of the source brightness. 
The second process is required to ensure that each data set is fairly weighted 
according to error bars. We eliminate $3\sigma$ outliers in the analysis 
in order to minimize their effect on modeling.

\begin{deluxetable*}{lrrrrrrr}
\tablecaption{Lensing Parameters\label{table:one}}
\tablewidth{0pt}
\tablehead{
\multicolumn{1}{l}{Parameters} &
\multicolumn{1}{c}{Standard} &
\multicolumn{2}{c}{Parallax} &
\multicolumn{2}{c}{Orbit} &
\multicolumn{2}{c}{Orbit+Parallax} \\
\multicolumn{1}{c}{} &
\multicolumn{1}{c}{} &
\multicolumn{1}{c}{$u_0>0$} &
\multicolumn{1}{c}{$u_0<0$} &
\multicolumn{1}{c}{$u_0>0$} &
\multicolumn{1}{c}{$u_0<0$} &
\multicolumn{1}{c}{$u_0>0$} &
\multicolumn{1}{c}{$u_0<0$}
}
\startdata
$\chi^2$/dof              & 13472.0/13396       & 13428.8/13394       & 13433.9/13394       & 13459.8/13394       & 13458.6/13394       & 13405.0/13392       & 13422.0/13392       \\
$t_0$ (${\rm HJD'}$)      & 6406.71 $\pm$ 0.17  & 6406.47 $\pm$ 0.19  & 6406.38 $\pm$ 0.16  & 6405.54 $\pm$ 0.18  & 6405.51 $\pm$ 0.18  & 6406.18 $\pm$ 0.20  & 6406.30 $\pm$ 0.17  \\
$u_0$                     & 0.809   $\pm$ 0.010 &  0.832  $\pm$ 0.011 & -0.827  $\pm$ 0.006 &  0.871  $\pm$ 0.008 & -0.875  $\pm$ 0.009 &  0.811  $\pm$ 0.011 & -0.812  $\pm$ 0.008 \\
$t_{\rm E}$ (days)        & 38.5    $\pm$ 0.3   &  38.9   $\pm$ 0.4   &  39.3   $\pm$ 0.3   &  38.3   $\pm$ 0.3   &  38.2   $\pm$ 0.3   &  37.6   $\pm$ 0.4   &  37.1   $\pm$ 0.4   \\
$s$                       & 0.594   $\pm$ 0.003 &  0.595  $\pm$ 0.003 &  0.596  $\pm$ 0.002 &  0.575  $\pm$ 0.002 &  0.573  $\pm$ 0.003 &  0.607  $\pm$ 0.004 &  0.607  $\pm$ 0.002 \\
$q$                       & 0.146   $\pm$ 0.005 &  0.112  $\pm$ 0.005 &  0.109  $\pm$ 0.003 &  0.072  $\pm$ 0.002 &  0.067  $\pm$ 0.002 &  0.130  $\pm$ 0.007 &  0.143  $\pm$ 0.002 \\
$\alpha$ (rad)            & 6.36    $\pm$ 0.02  &  6.24   $\pm$ 0.02  & -6.22   $\pm$ 0.01  &  6.09   $\pm$ 0.01  & -6.07   $\pm$ 0.01  &  6.36   $\pm$ 0.02  & -6.39   $\pm$ 0.01  \\
$\rho_\ast$ ($10^{-3}$)   & 10.28   $\pm$ 0.21  & 10.27   $\pm$ 0.30  &  9.91   $\pm$ 0.21  &  8.16   $\pm$ 0.24  &  7.85   $\pm$ 0.16  & 11.12   $\pm$ 0.24  & 11.71   $\pm$ 0.33  \\
$\pi_{{\rm E}, N}$        & --                  &  0.11   $\pm$ 0.07  & -0.01   $\pm$ 0.05  &  --                 &  --                 &  0.48   $\pm$ 0.05  & -0.44   $\pm$ 0.07  \\
$\pi_{{\rm E}, E}$        & --                  & -0.17   $\pm$ 0.03  & -0.18   $\pm$ 0.03  &  --                 &  --                 & -0.06   $\pm$ 0.03  & -0.20   $\pm$ 0.03  \\
$ds/dt$ (yr$^{-1}$)       & --                  & --                  & --                  &  0.44   $\pm$ 0.05  &  0.51   $\pm$ 0.03  & -0.31   $\pm$ 0.05  & -0.48   $\pm$ 0.06  \\
$d\alpha/dt$ (yr$^{-1}$)  & --                  & --                  & --                  & -1.48   $\pm$ 0.07  &  1.64   $\pm$ 0.04  & -0.46   $\pm$ 0.18  &  0.05   $\pm$ 0.13
\enddata 
\vspace{0.05cm}
\tablecomments{
${\rm HJD}'= {\rm HJD}-2450000$
}
\end{deluxetable*}

\begin{figure}[th]
\epsscale{1.15}
\plotone{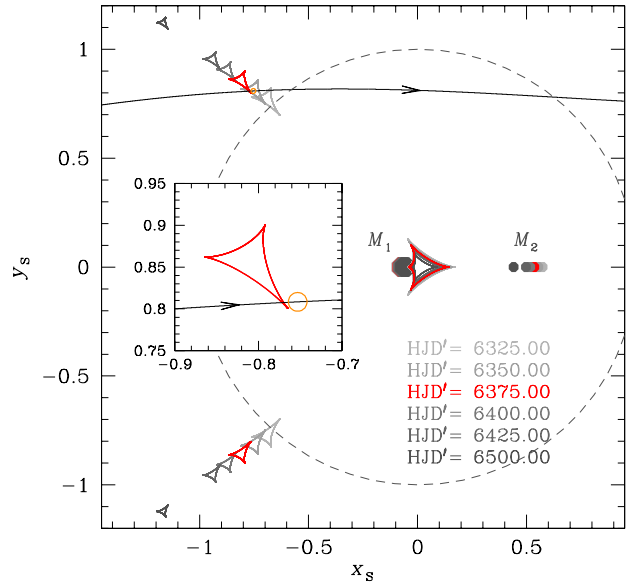}
\caption{\label{fig:two}
Geometry of the source trajectory (curve with an arrow) 
with respect to the lens components ($M_{1}$ and $M_{2}$) 
and caustics (closed figures composed of concave curves). 
The dashed circle represents the Einstein ring. 
The coordinates are centered at the barycenter of the binary lens 
and all lengths are scaled to the Einstein radius 
corresponding to the total mass of the binary. 
Caustics vary in time due to the lens orbital motion. 
The red caustics correspond to the time of the anomaly. 
The inset shows the enlarged view of the source star's caustic crossing. 
The empty orange circle represents the source size relative to the caustic. 
}
\end{figure}

\begin{figure}[th]
\epsscale{1.1}
\plotone{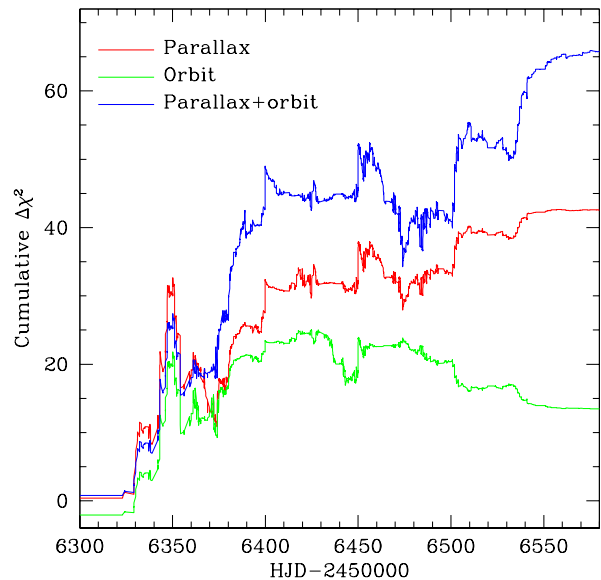}
\caption{\label{fig:three}
Cumulative distributions of $\Delta\chi^{2}$ between the models 
considering higher-order effects relative to the standard model as a function of time. 
}
\end{figure}

\section{Analysis}

Under the approximation that the relative lens-source motion is rectilinear, 
the light curve of a binary-lens event is described by 7 standard parameters. 
See Appendix A for graphical presentation of the parameters. 
In our modeling of the light curve, 
we use the center of mass of the binary lens as the reference position.

Searching for the set of lensing parameters 
that best describes the observed light curve is done through multiple stages. 
At the first stage, we explore $\chi^{2}$ surface in the parameter space and 
locate all possible local minima by conducting a grid search 
for a subset of the lensing parameters. At the second stage, 
we further refine each local minimum. At the last stage, we identify the global minimum 
by comparing $\chi^{2}$ values of the individual local solutions. 
Grid search at the first stage is done in the parameter space of $(s,q,\alpha)$ 
for which the lensing magnifications can vary dramatically 
with small changes of the parameters. For the other parameters, 
for which lensing magnifications vary smoothly for the changes of the parameters, 
we search for solutions by minimizing $\chi^{2}$ using a  Markov Chain Monte Carlo (MCMC). 
Once a solution is found, the uncertainty of each lensing parameter is estimated 
based on the distribution of parameters derived from the corresponding MCMC chain.

To model the anomaly, which was produced by the caustic crossing of the source star, 
it is needed to consider finite-source effects. 
For the computation of finite magnifications, 
we use the numerical method of the inverse ray-shooting method \citep{kayser86,schneider87} 
for the central region of the perturbation and semi-analytic 
hexadecapole approximation \citep{pejcha09,gould08} 
for the vicinity of the perturbation. 
We account for the surface-brightness variation of the source 
caused by limb darkening by modeling the surface-brightness profile 
using a linear limb-darkening law. We adopt the limb-darkening coefficients 
$(u_V, u_R, u_I) = (0.81, 0.73, 0.63)$ from \citet{claret00} 
using $T_{\rm eff} = 4500~{\rm K}$ and 
${\rm log}~g = 2$, where $T_{\rm eff}$ and ${\rm log}~g$ are derived 
from the de-reddened brightness and color of the source star (see Section 4). 
For the MOA data taken by using a non-standard filter system, 
we use $u_{RI} = (u_R+u_I)/2 = 0.68$.

For some binary lensing events, the seven basic parameters are not adequate to 
precisely describe lensing light curves. 
These cases often occur for long time-scale events 
where the assumption of the rectilinear lens-source motion is no longer valid. 
Parallax effects occur due to the change of the observer's position 
caused by the orbital motion of the Earth around the Sun \citep{gould92,alcock95}, 
causing non-rectilinear source motion. 
Similarly, lens orbital effects also cause non-rectilinear source motion 
due to the change of the lens positions caused by the orbital motion of the lens 
\citep{dominik98,albrow00,shin11,jung13,park13}.
Parallax effects are described by two parameters $\pi_{{\rm E}, N}$ and $\pi_{{\rm E}, E}$ 
which are the two components of the lens parallax vector ${\pivec}_{\rm E}$ 
projected onto the sky along the north and east equatorial coordinates, respectively. 
The magnitude of ${\pivec}_{\rm E}$ corresponds to 
the ratio of the lens-source relative parallax $\pi_{\rm rel}$ 
to the angular Einstein radius $\theta_{\rm E}$, i.e.
\begin{equation}
{\pi}_{\rm E}={\pi_{\rm rel} \over \theta_{\rm E}};\qquad  
\pi_{\rm rel}={{\rm AU}\left({1 \over D_{\rm L}} - {1 \over D_{\rm S}}\right)}, 
\label{eq1}
\end{equation} 
where $D_{\rm L}$ and $D_{\rm S}$ are 
the distances to the lens and source, respectively \citep{gould04}. 
The direction of ${\pivec}_{\rm E}$ corresponds to the lens-source relative motion.
To the first-order approximation, 
lens-orbital effects are described by two parameters 
$ds/dt$ and $d\alpha/dt$ that represent the change rates of 
the projected binary separation and the source-trajectory angle, respectively.

\begin{figure}[ht]
\epsscale{1.15}
\plotone{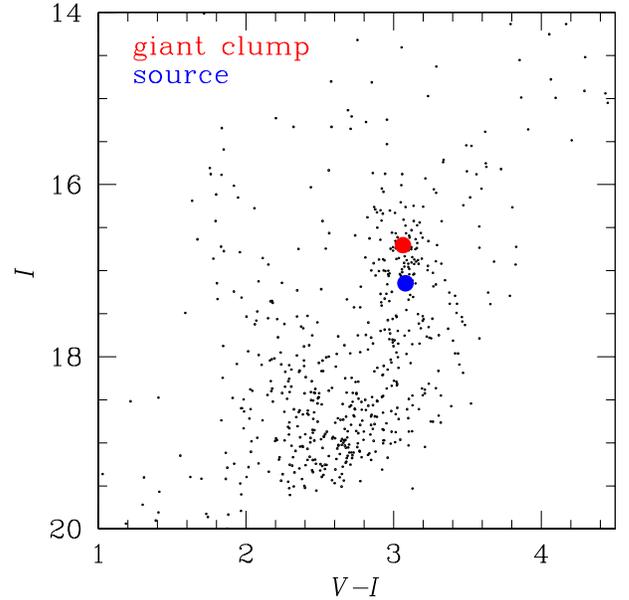}
\caption{\label{fig:four}
Instrumental color-magnitude diagram of stars in the region around the lensed star. 
The locations of the lensed star (source) and the centroid of giant clump are marked.
}
\end{figure}

\begin{figure*}[th]
\epsscale{0.9}
\plotone{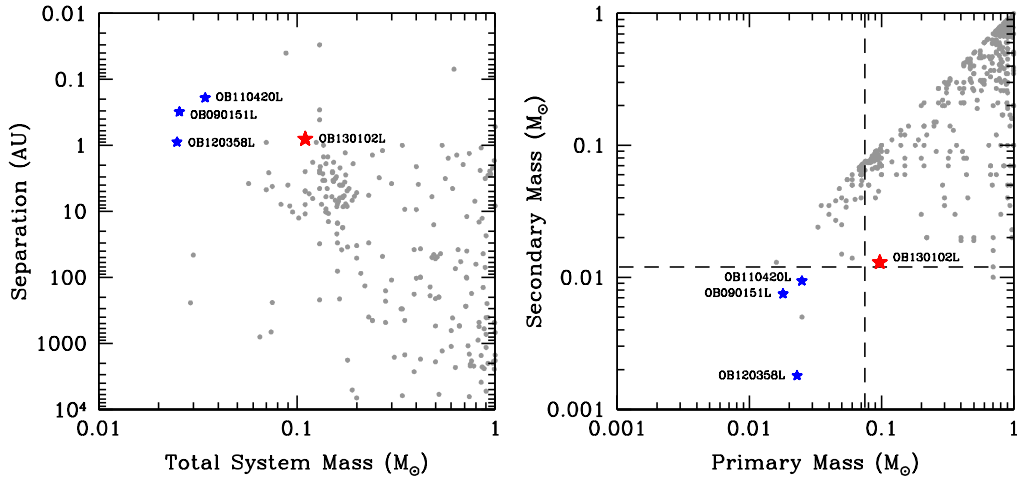}
\caption{\label{fig:five}
Total mass vs. separation (left panel) and primary vs. secondary masses (right panel) 
for a compilation of low-mass binaries. 
Microlensing binaries are denoted in `star' marks while 
those discovered by other methods are marked by dots. 
The red star is the microlensing binary reported in this work and the 
three blue stars are the binaries reported by \citet{choi13} and \citet{han13}. 
The vertical and horizontal dashed lines 
represent the star/brown-dwarf and brown-dwarf/planet boundaries, respectively.
}
\end{figure*}

Measurements of higher-order effects are important 
for the determination of the physical lens parameters. 
By measuring the normalized source radius $\rho_{\ast}$ 
from the analysis of the light curve affected by 
finite-source effects, one can measure the Einstein radius by
\begin{equation}
\theta_{\rm E} = {\theta_{\ast} \over \rho_{\ast}},
\label{eq2}
\end{equation}
where the angular source radius $\theta_{\ast}$ is derived from 
the information about the source star (see Section 4). 
Then, along with the lens parallax measured from the analysis of 
the long-term deviation caused by parallax effects, 
the mass and distance to the lens are determined by
\begin{equation}
M_{\rm tot}={\theta_{\rm E} \over \kappa\pi_{\rm E}};\qquad  
D_{\rm L}={{\rm AU}\over \pi_{\rm E}\theta_{\rm E}+\pi_{\rm S}},
\label{eq3}
\end{equation}
respectively. 
Here $\kappa=4G/(c^{2}{\rm AU})$ and $\pi_{\rm S}={\rm AU}/{D_{\rm S}}$ 
is the parallax of the source star.

We test various models considering the individual and 
combinations of the higher-order effects. 
In the ``standard'' model, the light curve is fitted based on the 
7 standard lensing parameters. In the ``parallax'' and ``orbit'' models, 
we separately consider the parallax and lens-orbital effects. 
Finally, in the ``orbit+parallax'' model, 
we consider both the parallax and orbital effects. 
When the higher-order effects are considered, 
we test two solutions resulting from 
``ecliptic degeneracy'' \citep{skowron11}. 
The two solutions resulting from this degeneracy 
have almost identical parameters except 
$(u_0,~\alpha,~\pi_{{\rm E}, N},~d\alpha/dt) \rightarrow -(u_0,~\alpha,~\pi_{{\rm E}, N},~d\alpha/dt)$.

\section{Physical Parameters}

In Table~\ref{table:one}, we list the results of analysis for the models that we tested. 
The model light curve of the best-fit solution (orbit+parallax with $u_0 > 0$) 
is superposed on the data in Figure~\ref{fig:one}. 
Figure~\ref{fig:two} shows the geometry of the lens system where the source trajectory 
with respect to the lens positions and caustics are presented. 
It is found that the event was produced by a binary with a mass ratio 
between the components is $q = 0.13$. 
The projected separation between the binary components is $s = 0.61$, 
which is less than the Einstein radius. In this case, the caustic is composed of 
three closed curves where one is located near the barycenter of the binary lens 
and the other two are located away from the central region. 
The anomaly was produced by the passage of the source trajectory 
over one of the outer caustics.

We find that the higher-order effects improve the fit. 
As measured by $\chi^{2}$ difference from the standard model, 
the fit improvements are $\Delta\chi^{2} = 43.2$ and $13.4$ 
when the parallax and orbital effects are separately considered. 
When both effects are simultaneously considered, 
the improvement is $\Delta\chi^{2} = 67.0$, 
which is $> 8\sigma $. While this is formally significant, 
careful diagnosis of the signal is needed 
because in microlensing subtle systematic trends might masquerade as signals. 
We therefore check the possibility of systematics by inspecting 
where the signal of the higher-order effects comes from. 
If systematics in data affected the fit, 
the signal would come from localized epochs of the event. 
By contrast, if the signal is due to genuine higher-order effects, 
it would come throughout the event 
because both the orbital motions of the Earth and the lens 
have long-term effects on the lensing light curve. 
In Figure~\ref{fig:three}, 
we present the cumulative distribution of $\Delta\chi^{2}$ as a function of time. 
Although there exist several local fluctuations and thus 
possibility of systematics cannot be completely ruled out, 
$\chi^{2}$ improvement occurs throughout the event, 
suggesting that the signals of higher-order effects are real.  
Since the anomaly is well covered, 
finite-source effects are clearly detected, yielding a normalized source radius 
$\rho_{\ast} = (11.1 \pm 0.2)\times10^{-3}$.

\begin{deluxetable}{lrr}[h]
\tablecaption{Physical Parameters\label{table:two}}
\tablewidth{0pt}
\tablehead{
\multicolumn{1}{c}{Parameters} &
\multicolumn{1}{c}{$u_0>0$} &
\multicolumn{1}{c}{$u_0<0$} 
}
\startdata
Angular Einstein radius (mas)                  &  0.43  $\pm$ 0.04  &  0.41  $\pm$ 0.03     \\
Geocentric proper motion (mas $\rm yr^{-1}$)   &  4.19  $\pm$ 0.37  &  4.06  $\pm$ 0.34     \\
Heliocentric proper motion (mas $\rm yr^{-1}$) &  4.30  $\pm$ 0.38  &  3.69  $\pm$ 0.31     \\
Total mass ($M_\odot$)                         &  0.108 $\pm$ 0.014 &  0.104 $\pm$ 0.017    \\
Primary mass ($M_\odot$)                       &  0.096 $\pm$ 0.013 &  0.091 $\pm$ 0.015    \\
Companion mass ($M_\odot$)                     &  0.012 $\pm$ 0.002 &  0.013 $\pm$ 0.002    \\
Distance (kpc)                                 &  3.04  $\pm$ 0.31  &  3.15  $\pm$ 0.37     \\ 
Projected separation (AU)                      &  0.80  $\pm$ 0.08  &  0.79  $\pm$ 0.09     \\ 
$({\rm KE}/{\rm PE})_{\perp}$                  &  0.028 $\pm$ 0.006 &  0.037 $\pm$ 0.005
\enddata  
\vspace{0.05cm}
\end{deluxetable}

With both finite-source and parallax effects measured, the mass and distance 
to the lens are estimated by using the relations in Equations (\ref{eq3}).  
For this, we estimate the angular source radius $\theta_{\ast}$ based on the 
color and brightness.  The angular Einstein radius is estimated following two 
steps.  In the first step, we estimate the de-reddened color $(V-I)_{\rm 0}$ 
and the brightness $I_{\rm 0}$ of the source star by using the centroid of Bulge 
clump giants for which its de-reddened color, $(V-I)_{\rm 0, c} = 1.06$, 
and brightness, $I_{\rm 0,c} = 14.45$, are known from independent measurements 
\citep{bensby13,nataf13}.  This method is valid because the source and Bulge 
giants are at nearly same distances and thus experience almost same 
extinction \citep{yoo04}.  The estimated color and brightness of the source star 
are $(V-I, I)_{\rm 0} = (1.08, 14.97)$.  Figure~\ref{fig:four} shows the location 
of the source star in the color-magnitude of neighboring stars with respect to the 
centroid of clump giants, indicating that the source is a K-type giant.  
In the second step, we convert $V-I$ into $V-K$ by using the color-color 
relation \citep{bessell88}.  Then the angular source radius is estimated 
by adopting the relation between $V-K$ and $\theta_{\ast}$ provided 
by \citet{kervella04}.  The derived angular source radius is 
$\theta_{\ast} = 4.80 \pm 0.41$ ${\mu}$as.  The error on $\theta_{\ast}$ 
comes from three major sources: (1) the uncertainty in the source flux $f_{\rm S}$, 
(2) the uncertainty involved with the conversion from color to $\theta_{\ast}$, 
and (3) other auxiliary uncertainties concerned with processes such as 
positioning the centroid of giant stars, $V-I$ to $V-K$ conversion, etc. 
We estimate that the uncertainty in $f_{\rm S}$ is $\sigma_{f_{\rm S}} = 5\%$.  
Uncertainties of microlensing colours $\sigma_{(V-I)_{\rm 0}} = 0.07$ 
mag \citep{bensby13} and brightness $\sigma_{I_{\rm 0, \rm c}} = 0.09$ 
mag \citep{nataf13} of clump giant stars contribute $\sim6\%$ error 
in $\theta_{\ast}$ measurement. By considering other factors, we adopt 
the uncertainty from factors (2) and (3) as $7\%$. The estimated source radius 
corresponds to the angular Einstein radius $\theta_{\rm E} = 0.43 \pm 0.04$ mas.

\begin{figure}[t]
\epsscale{1.0}
\plotone{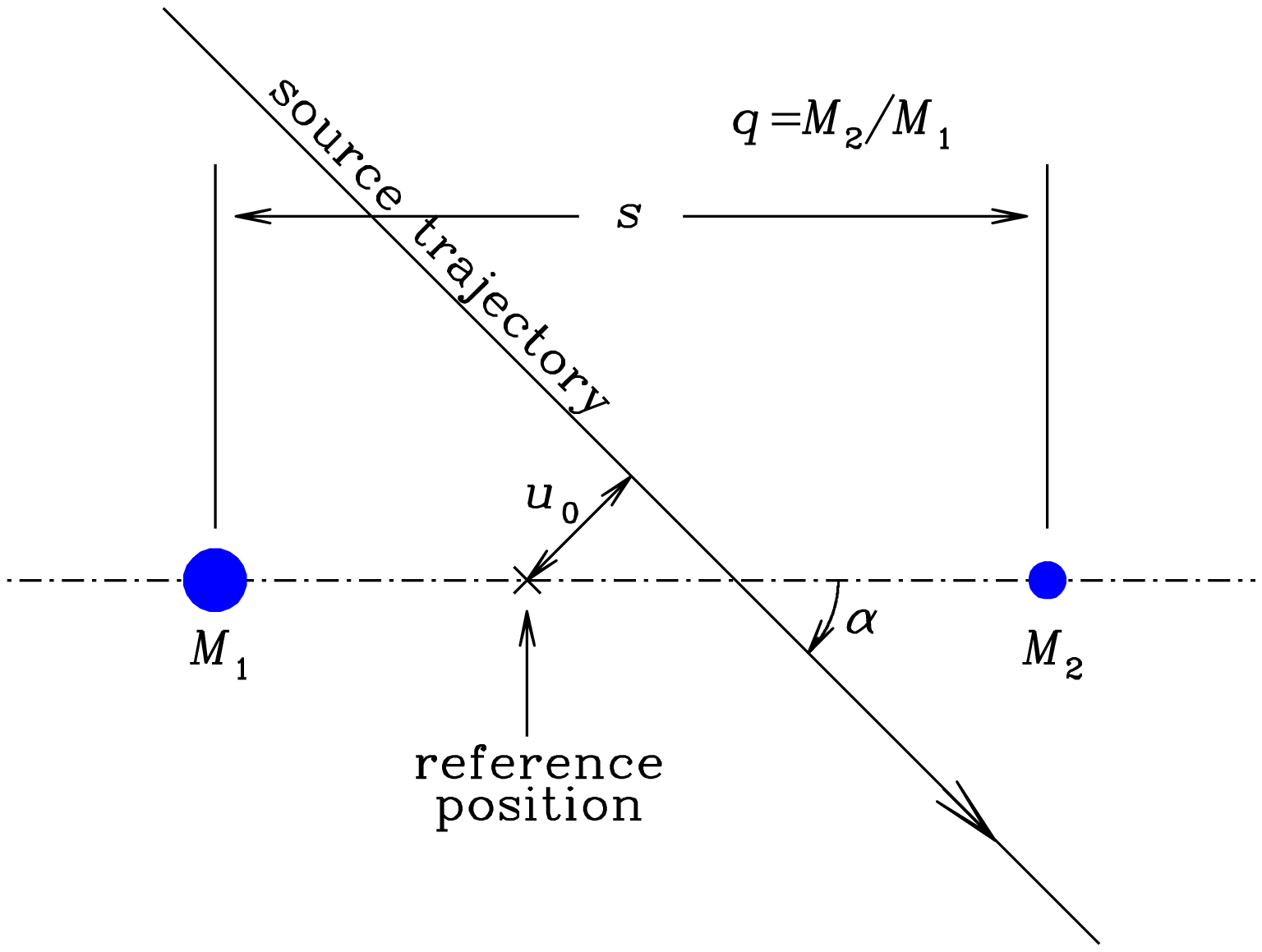}
\caption{\label{fig:six}
Graphical presentation of binary-lensing parameters. 
The filled dots marked by $M_1$ and $M_2$ 
represent the locations of the lens components. 
The straight line with an arrow is the source trajectory. 
}
\end{figure}

In Table~\ref{table:two}, we summarize the determined physical quantities
of the lens system.  We present 2 sets of quantities corresponding
to the $u_0 > 0$ and $u_0 < 0$ solutions resulting from
the ecliptic degeneracy.  It is found that the $u_0 > 0$ model
is preferred by $\Delta\chi^{2} = 17.0$.
This is formally $> 4\sigma$ level, but one cannot completely 
rule out $u_0 <0$ model considering possible systematics in data.
However, we note that both solutions result
in similar physical quantities. According to the best-fit model, the
masses of the lens components are $M_{1} = 0.096 \pm 0.013~M_{\odot}$ and 
$M_{2} = 0.012 \pm 0.002~M_{\odot}$. The distance to the lens is
$D_{\rm L} = 3.04 \pm 0.31$ kpc and the projected separation between the
lens components is $r_\perp = sD_{\rm L}\theta_{\rm E}=0.80\pm 0.08$ AU. 
In order to check the validity of the solution, we also present 
the {\it projected} kinetic to potential energy ratio, which is computed by
\begin{equation}
\left( {\rm KE \over \rm PE} \right)_{\bot} = {(r_{\bot}/{\rm AU})^{2} \over 8{\pi}^{2}(M_{\rm tot}/M_{\odot})}
\left[ \left( {1 \over s}{ds \over dt} \right)^{2} + \left( { d\alpha \over dt} \right)^{2} \right],
\label{eq4}
\end{equation}
where $M_{\rm tot}$ is the total mass of the binary-lens \citep{dong09}. 
We note that the lensing parameters $ds/dt$ and $d\alpha/dt$ 
are determined from modeling considering the orbital motion of the lens. 
The ratio should be less than unity to be a bound system. The measured ratio is 
$({\rm KE}/{\rm PE})_{\perp} < 1.0$ and thus meets the condition of boundness. 
Its small value tends to imply that the true separation is 
several times larger than the projected separation.

In Figure~\ref{fig:five}, we compare the physical parameters of OGLE-2013-BLG-0102L
to those of low-mass binaries from the VLM binaries archive (http://www.vlmbinaries.org) 
and other references 
\citep{basri99,lane01,burgasser08,faherty11,burgasser12}. Also 
marked are the three low-mass microlensing binaries: OGLE-2009-BLG-151L and 
OGLE-2011-BLG-420L reported by \citet{choi13} and 
OGLE-2012-BLG-0358L reported by \citet{han13}.
It is found that the microlensing 
binaries are located in the low-mass, close-separation, and low-mass-ratio regions 
in the parameter space. Among known VLM binaries, 
we find that only `Cha H$\alpha$ 8' has similar physical parameters: 
$r_{\bot} = 1.3~{\rm AU}$, $M_{1} = 0.1~M_{\odot}$, 
$M_{2} = 0.019~M_{\odot}$ \citep{joergens06,joergens07}. 
However, this binary was discovered in a star-forming cloud and thus very young. 
Therefore, the reported VLM binary demonstrates that microlensing provide 
an important method that can complement other methods.

The discovered binary is of scientific interest 
because the masses of the primary and the companion 
correspond to the upper and lower limits of brown dwarfs, respectively. 
Although there exist some dispute, the popular convention for the division 
between low-mass stars and brown dwarfs is $\sim 0.075~M_{\odot}$, 
below which hydrogen fusion reaction in cores does not occur \citep{burrows97}, 
while the convention for the division between brown dwarfs and giant planets is 
$\sim 0.012~M_{\odot}~(13~M_{\rm J})$, below which deuterium burning cannot 
be ignited \citep{spiegel11}. Then, the individual binary components of 
OGLE-2013-BLG-0102L have masses near the hydrogen-burning and 
deuterium-burning mass limits, respectively.

\section{Conclusion}

We found a very low-mass binary from the observation and analysis of the 
microlensing event OGLE-2013-BLG-0102. 
The event was characterized by a strong short-term anomaly superposed on 
a smoothly varying lensing curve with a moderate magnification.
It was found that the event was produced by a binary object with a mass 
ratio between the components is $q = 0.13$ and the anomaly was caused by 
the passage of the source trajectory over a caustic located away from 
the barycenter of the binary lens. By measuring deviations in lensing light curve 
caused by both finite-source and parallax effects,
we determined the physical parameters of the lens.  
It was found that the lens is composed of objects with masses 
 $M_{1} = 0.096 \pm 0.013~M_{\odot}$ and 
$M_{2} = 0.012 \pm 0.002~M_{\odot}$, which correspond to 
the hydrogen-burning and deuterium-burning limits, respectively. 
The binary is located at a distance $3.04 \pm 0.31~{\rm kpc}$ and 
the projected separation between the components is 
$0.80 \pm 0.08~{\rm AU}$.

\acknowledgments
Work by C.H. was supported by Creative Research Initiative Program
(2009-0081561) of National Research Foundation of Korea. A.G. and 
B.S.G. acknowledge support from NSF AST-1103471. B.S.G., A.G., and 
R.W.P. acknowledge support from NASA grant NNX12AB99G. The OGLE project has received 
funding from the European Research Council under the European Community's 
Seventh Framework Programme (FP7/2007-2013)/ERC grant agreement No. 
246678 to A.U. The MOA experiment was supported by grants JSPS22403003 
and JSPS23340064. T.S. acknowledges the support JSPS 24253004. T.S. 
is supported by the grant JSPS23340044. Y.M. acknowledges support 
from JSPS grants JSPS23540339 and JSPS19340058. 
Work by J.C.Y. was performed in part under contract 
with the California Institute of Technology (Caltech) funded by NASA 
through the Sagan Fellowship Program.

\appendix
\section{BINARY-LENSING PARAMETERS}

\smallskip 

For the basic description of binary-lens events, 7 lensing parameters are needed. 
Three of these parameters describe the source-lens approach: 
the time of the closest source approach to a reference position of the lens, $t_0$, 
the separation between the source and the reference position at $t_0$, $u_0$ (impact parameter), 
and the time scale for the source to cross the Einstein radius 
corresponding to the total mass of the binary lens, $t_{\rm E}$ (Einstein time scale). 
Another three parameters describe the binary lens: the mass ratio, $q = M_2/M_1$, 
the projected separation between the lens components, $s$, 
and the angle between the source trajectory and the binary axis, $\alpha$ (source trajectory angle). 
See the graphical presentation of the parameters in Figure~\ref{fig:six}. 
We note that parameters $u_0$ and $s$ are normalized to the angular Einstein radius $\theta_{\rm E}$. 
The last parameter is the source radius normalized to the Einstein radius, 
$\rho_{\ast}$ (normalized source radius).

\end{document}